\documentclass[12pt]{iopart}

\usepackage{epsfig}

\usepackage{iopams}  

\newcommand{\bemm}{\begin{multline}}
\newcommand{\enm}{\end{multline}}
\newcommand{\beq}{\begin{equation}}
\newcommand{\eeq}{\end{equation}}
\newcommand{\beqa}{\begin{eqnarray}}
\newcommand{\eeqa}{\end{eqnarray}}
\newcommand{\ba}{\begin{array}}
\newcommand{\ea}{\end{array}}

\newtheorem{teo}{Theorem}

\newcommand{\be}{\begin{equation}}

\newcommand{\ee}{\end{equation}}
\newcommand{\bt}{\begin{teo}}
\newcommand{\et}{\end{teo}}

\newcommand{\om}{\omega}

\newcommand{\on}{\omega_0}
\newcommand{\ok}{\omega_1}

\begin{document}

\epsfig{file=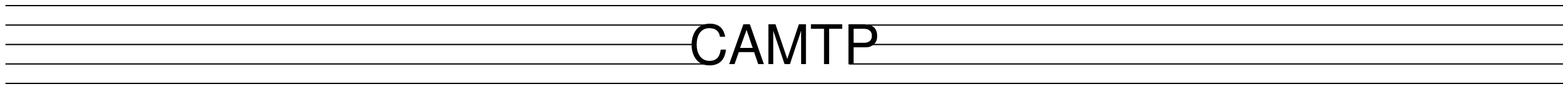,height=8mm,width=\textwidth}

\letter{Exact energy distribution function in time-dependent
harmonic oscillator}

\author{ Marko Robnik$^1$, Valery G. Romanovski$^1$ 
and Hans-J\"urgen St\"ockmann$^2$ }

\address{$^1$CAMTP - Center for Applied Mathematics and Theoretical Physics,
University of Maribor, Krekova 2, SI-2000 Maribor, Slovenia}
\address{$^2$Fachbereich Physik, Philipps-Universit\"at Marburg, 
Renthof 5, D-35032 Marburg, Germany}

\eads{\mailto{Robnik@uni-mb.si}, \mailto{Valery.Romanovsky@uni-mb.si},
\mailto{Stoeckmann@physik.uni-marburg.de}}

\begin{abstract}
Following a recent work by Robnik and Romanovski (J.Phys.A: Math.Gen.
{\bf 39} (2006) L35, Open Syst. \& Infor. Dyn. {\bf 13} (2006) 197-222)
we derive the explicit formula for the universal distribution function 
of the final
energies in a time-dependent 1D harmonic oscillator, whose functional
form does not depend on the details of the frequency $\omega (t)$,
and is closely related to the conservation of the adiabatic invariant.
The normalized 
distribution function is $P(x) = \pi^{-1} (2\mu^2 - x^2)^{-\frac{1}{2}}$,
where  $x=E_1- \bar{E_1}$, $E_1$ is the final energy, $\bar{E_1}$ is its
average value, and $\mu^2$ is the variance of $E_1$. $\bar{E_1}$ and 
$\mu^2$ can be calculated exactly using the WKB approach to all orders.
\end{abstract}

\vspace{2pc}

In a recent work \cite{RR2006a,RR2006b} Robnik and Romanovski have
studied the energy evolution in a general 1D time-dependent
harmonic oscillator and studied the closely related questions of
the conservation of adiabatic invariants \cite{Einstein,Rob,LL,Rein,Henrard}. 
Starting with the ensemble
of uniformly distributed (w.r.t. the canonical angle variable) initial 
conditions on the initial invariant torus of energy $E_0$, they have calculated
the average final energy $\bar{E_1}$, the variance $\mu^2$ and all
higher moments. The even moments are powers of $\mu^2$, whilst the odd
moments are exactly zero, because the distribution function $P(E_1)$
of the final energies $E_1$ is an even function w.r.t. $\bar{E_1}$.
In this Letter we derive explicit formula for $P(E_1)$, namely
we shall derive 

\be \label{Main}
P(E_1) = {\rm Re} \frac{1}{\pi \sqrt{2\mu^2 - x^2}} ,
\ee
where $x=E_1- \bar{E_1}$, and Re denotes the real part,
so that (\ref{Main}) is zero
for $|x| > \mu\sqrt{2}$. This we do by using the exact results
for the higher (even) moments and by employing the characteristic
function $f(y)$ of $P(x)$.

The dynamics of our system is described by the Newton equation  

\be \label{Newton}
\ddot{q} + \omega^2 (t) q =0
\ee
which is generated by the system's  Hamilton function 
$H = H(q,p,t)$, whose numerical value $E(t)$ at time $t$ is precisely the total
energy of the system at time $t$, and in case of 1D harmonic oscillator this is

\be \label{Hamilton}
H=\frac {p^2}{2M}+\frac 12 M \omega^2(t)q^2,
\ee
where $q,p,M,\omega$ are the coordinate, the momentum, the mass and
the frequency of the linear oscillator, respectively. 

The dynamics is linear in $q,p$, as described by (\ref{Newton}),
but nonlinear as a function of $\omega (t)$ and therefore is subject to
the nonlinear dynamical analysis. By using the index $0$ and $1$ we denote the
initial ($t=t_0$) and final $(t=t_1)$ value of the variables.

The transition map $\Phi$ maps initial conditions $(q_0,p_0)$
onto the final conditions $(q_1,p_1)$
\be
\Phi:\left( 
\begin{array}{c}
q_0\\ p_0
\end{array}
 \right)\mapsto  \left( 
\begin{array}{c}q_1\\ p_1
\end{array}
 \right) = \left(
\begin{array}{cc}
 a & b\\
  c & d
\end{array}
\right) \left( \begin{array}{c}
q_0\\ p_0
\end{array}
 \right)
\ee
with  $\det(\Phi)=ad-bc=1$, and $a,b,c,d$ can be calculated as
shown in \cite{RR2006a,RR2006b}.
Let $E_0=H(q_0,p_0,t=t_0)$ be the initial energy and 
$E_1=H(q_1,p_1,t=t_1)$ be the final energy.
Introducing the new coordinates, namely the action $I=E/\omega$ and
the angle $\phi$, and assuming the uniform distribution of initial
angles $\phi$ over the period $2\pi$, we can immediately calculate
the final average energy $\bar{E_1}$ and  the variance 

\be \label{m2}
\mu^2 = 
\overline{(E_1-\bar E_1)^2}=\frac {E_0^2}2 \left[\left( \frac{\bar E_1}{E_0} 
 \right)^2 - \frac{\om_1^2}{\om_0^2} \right].
\ee
It is shown in reference \cite{RR2006b} that 
for arbitrary positive integer $m$, we have  
$\overline{(E_{1}-\bar E_{1})^{2m-1}}=0$ and

\be\label{emeven}
\overline{(E_{1}-\bar E_{1})^{2m}}=\frac {(2m -1)!!}{ m!}
\left( \overline{(E_{1}-\bar E_{1})^2}\right)^m.
\ee
Thus $2m$-th moment of $P(E_1)$ is equal to $(2m-1)!! \mu^{2m}/m!$,
and therefore, indeed, all moments of $P(E_1)$ are uniquely determined 
by the first moment $\bar{E_1}$. Obviously, $P(E_1)$ is in this sense
universal, because it depends only on the average final energy $\bar{E_1}$ and
the ratio $\ok/\on$ of the final and initial frequencies, and 
does not depend otherwise on any details of $\om (t)$. It has a finite
support  $(E_{min},E_{max})= (\bar{E_1}-\mu\sqrt{2}, \bar{E_1}+\mu\sqrt{2})$, 
it is an even distribution w.r.t.
$\bar{E_1}=(E_{min}+E_{max})/2$, and has an integrable singularity
of the type $1/\sqrt{x}$ at both $E_{min}$ and $E_{max}$. This singularity
stems from a projection of the final ensemble at $t_1$ onto the curves of
constant final energies $E_1$ of $H(q,p,t_1)$. Of course, all that we say
here for the distribution of energies $E_1$ holds true also for the 
final action, the adiabatic invariant  $I_1=E_1/\om_1$. It is perhaps
worthwhile to mention that the moments of our distribution according to
(\ref{emeven}) grow as $2^m/\sqrt{\pi m}$, whilst e.g. in the Gaussian
distribution they grow much faster, namely as $2^m\Gamma (m+1/2)/\sqrt{\pi}$,
where $\Gamma (x)$ denotes the gamma function.

Now we derive the distribution function (\ref{Main}) using the
characteristic function $f(y)$ of $P(x)$, namely

\be \label{Charac}
f(y) = \int_{-\infty}^{\infty} e^{iyx} P(x) dx.
\ee
We see immediately that the $n$-th derivative at $y=0$ is equal to

\be \label{nder}
f^{(n)}(0) = \int_{-\infty}^{\infty} (ix)^n  P(x) dx = i^n \sigma_n,
\ee
where $\sigma_n$ is the $n$-th moment of $P(x)$ (in particular 
$\sigma_2 = \mu^2$), i.e.

\be \label{nmoment}
\sigma_n = \int_{-\infty}^{\infty} x^n P(x) dx.
\ee
Using the Taylor expansion for 
$f(y) = \sum_{n=0}^{\infty} \frac{f^{(n)}(0)}{n!} y^n$  and
the expressions  (\ref{emeven})  we see at once

\be \label{charac1}
f(y) = \sum_{m=0}^{\infty} \frac{i^{2m} \mu^{2m} (2m-1)!!}{m! (2m)!} y^{2m}
\ee
and using the formula  $(2m-1)!! = (2m)!/(2^m m!)$, we obtain,

\be \label{charac1}
f(y) = \sum_{m=0}^{\infty}  
\left(-\frac{\mu^2y^2}{2} \right)^m \frac{1}{(m!)^2},
\ee
which can be summed and is equal to the Bessel function \cite{GR}

\be \label{characf}
f(y) = J_0(\mu y \sqrt{2}).
\ee
Knowing the characteristic function (\ref{Charac},\ref{characf}) we
now only have to invert the Fourier transform,  namely

\be \label{invert}
P(x) = \frac{1}{2\pi} \int_{-\infty}^{\infty} e^{-iyx} f(y) dy =
\frac{1}{2\pi} \int_{-\infty}^{\infty} e^{-iyx} J_0(\sqrt{2} \mu y) dy,
\ee
which is precisely equal to (\ref{Main}) for $|x| \le \mu \sqrt{2}$
and is zero otherwise. (See reference \cite{GR}.) Indeed, the distribution
(\ref{Main}) is normalized to unity as it must be. It is essentially
the so-called $\beta (1/2,1/2)$  distribution or also termed
arc sine density \cite{Feller}, after shifting the
origin of $x$ to $1/2$ and rescaling of $x$. It is obvious that the 
distribution function does not depend on any details of $\omega (t)$
and is in this sense universal for 1D time-dependent harmonic oscillator
in case of {\em uniform canonical ensemble of initial conditions},
i.e. uniform w.r.t. the canonical angle.  

The rigorous method of calculating $\bar{E_1}$ and $\mu^2$ is explained
in \cite{RR2006a,RR2006b}. In the general case when (\ref{Newton})
cannot be solved exactly, the WKB method \cite{RR2000} 
can be applied to all orders,
as has been explained in detail in \cite{RR2006b}.

Let us now present the  direct algebraic 
derivation of the energy distribution function (\ref{Main}). 
By definition we have 

\be \label{PE1}
P(E_1)=
\frac 1{2\pi} \sum_{i=1}^4\left| \frac{d\phi}{dE_1} \right|_{\phi=\phi_j(E_1)},
\ee
where we have to sum up contributions from all four branches of the
function $\phi (E_1)$. Let us denote $x = E_1- \bar{E_1}$, so that
we have

\be \label{xeq1}
x = E_0 \left(\delta \cos (2\phi) + \gamma \sin (2\phi)\right)
= \mu \sqrt{2} \sin (2\phi + \psi),
\ee
where $\delta$ and $\gamma$ are expressed in terms of $a,b,c,d$ as
shown in \cite{RR2006a,RR2006b}, the variance  is
$\mu^2 = \frac{E_0^2}{2} (\delta^2 + \gamma^2)$, 
and $\tan \psi = \delta/\gamma$, so that
$\phi = \frac{1}{2} \arcsin \frac{x}{\mu \sqrt{2}} -\frac{\psi}{2}$.
Therefore  $|\frac{d\phi_i}{dx}| = \frac{1}{2\sqrt{2\mu^2 - x^2}}$ for
all four solutions $i=1,2,3,4$ and from (\ref{PE1}) we get 
(\ref{Main}) at once.

The first derivation demonstrates the power of
the approach employing the characteristic function $f(y)$ of $P(x)$,
giving us new insights,
whereas the second one leads elegantly to the final result,
eliminating many parameters by a rather elementary substitution.
In fact, this second method is geometrically obvious by the following
argument: In phase space, the initial distribution lies on an ellipse.
The final distribution lies on a different ellipse with the same area,
with the points equidistributed in the canonical angle
variable. This final ellipse intersects the energy contours
of the Hamiltonian, and thus after squeezing the final Hamiltonian
so that its energy contours are circles, the desired probability
distribution is simply the distribution of radii of those circles
through which the final ellipse passes.

In nonlinear systems the entire theory expounded in \cite{RR2006a,RR2006b}
must be reformulated and
as such it is an important open problem \cite{RR2006c}. 
For the case of a separatrix crossing some interesting numerical results 
have been obtained in \cite{RW}, namely $P(E_1)$ there 
has a substantial structure and is by far not so simple as (\ref{Main}). 

\ack

This   work was supported   by the Ministry of
Higher  Education, Science and Technology 
of the Republic of Slovenia,  Nova Kreditna Banka Maribor and 
 TELEKOM  Slovenije. We thank the referee for illuminating suggestions.

\end{document}